\documentclass[modern]{aastex631}

\begin{document}

\title{The Stellar Content of the Young Open Cluster Berkeley 50 (IC 1310)}

\author[0009-0006-1418-0702]{Meghan Speckert}
\affiliation{Lowell Observatory, 1400 W Mars Hill Road, Flagstaff, AZ 86001, USA}
\affiliation{Department of Astronomy and Planetary Science, Northern Arizona University, Flagstaff, AZ, 86011-6010, USA}
\email{mspeckert@lowell.edu}

\author[0000-0001-6563-7828]{Philip Massey}
\affiliation{Lowell Observatory, 1400 W Mars Hill Road, Flagstaff, AZ 86001, USA}
\affiliation{Department of Astronomy and Planetary Science, Northern Arizona University, Flagstaff, AZ, 86011-6010, USA}
\email{phil.massey@lowell.edu}

\author[0000-0001-5306-6220]{Brian A. Skiff}
\affiliation{Lowell Observatory, 1400 W Mars Hill Road, Flagstaff, AZ 86001, USA}
\email{bas@lowell.edu}


\begin{abstract}
We observed the Galactic open cluster Berkeley 50 in order to determine its stellar content, distance, and age. We obtained UBV photometry of 1145 stars in a 12\farcm3 $\times$ 12\farcm3  field, and used Gaia proper motions and parallaxes to identify 64 members, of which we obtained spectra of the 17 brightest members. The majority of the observed population we classified as B dwarfs, with the exception of a newly identified red supergiant star, which our spectroscopy shows has a B-type companion. Our study establishes the distance as 3.8~kpc, with an average color-excess $E(B-V)=0.9$.  Comparison of the physical properties of the cluster with the Geneva evolutionary tracks places the age of the cluster as 50-60~Myr, with its most massive members being $\sim7M_\odot$.

\end{abstract}


\section{Introduction} \label{sec:intro}

Berkeley 50 (IC 1310)\footnote{The cluster is located at $\alpha_{\rm 2000}$=20:10:00, $\delta_{\rm 2000}$=34:58. The first mention of this object, before it was assigned an IC designation, appears to be by \citet{1894MNRAS..54..327E}, where it is item 6 on his list of nebulous objects which did not appear in the New General Catalog.  He refers to it as a faint nebulosity, but admits that such objects may be a ``gathering of very faint stars." }
is a poorly studied Galactic open cluster. \citet{2008MNRAS.389..285T} included it in a catalog of previously unstudied Berkeley clusters.  Using near-IR photometry from 2MASS \citep{2003yCat.2246....0C}, and proper motions from the USNO
\citep{2004AAS...205.4815Z}, \citet{2008MNRAS.389..285T} determined an age of 250~Myr, a distance of 2.1$\pm$0.2 kpc, and a diameter of 7\arcmin.  Subsequently the cluster has been included in several statistical studies of Galactic open clusters as noted in the Unified Cluster Catalog \citep{2023MNRAS.526.4107P}, namely \citet{2012A&A...543A.156K}, \citet{2020A&A...640A...1C}, \citet{2021MNRAS.504..356D}, and \citet{2023A&A...673A.114H}.  The last three are based on Gaia data.  \citet{2023A&A...673A.114H} uses a statistical neural network approach to derive a much younger age, 35-89~Myr, and a much closer distance,  3.74-3.82~kpc.

Our interest in this cluster was piqued when one of us (BAS) noticed that one of its central stars (TYC 2679-0322-1, $V=11.9$ according to \citealt{2015AAS...22533616H}) is likely a previously overlooked red supergiant (RSG) based upon its 2MASS colors and brightness ($J-K=1.37$, $K=5.83$). The Data Release 3 (DR3) Gaia parallax for this red star corresponds to a likely distance of 5.1~kpc \citep{2021AJ....161..147B} but is flagged as uncertain \citep{gaiadr3}. We were able to obtain a spectrum of this star (as described below), which confirmed the star was of spectral type K5-7, and not, say, a strongly reddened F-type star. A back-of-the-envelope calculation showed that at distances of 2-5~kpc, and likely reddenings, the star's absolute magnitude would be consistent with it being a supergiant and not a giant. Of course, it could still be a chance superposition, but its location at the cluster's center was suggestive. If so, the age must be much smaller than the 250~Myr found by \citet{2008MNRAS.389..285T}, as even the oldest RSGs have ages of less than 70~Myr (see, e.g., \citealt{2012A&A...537A.146E}). We concluded that the stellar content of this cluster was worth investigating.

Even a casual inspection of the region shows that Berkeley 50 is not a populous cluster. (Before reviewing the proper motion data discussed below, we were not entirely convinced it actually was a cluster and not a chance supposition of several bright stars.)  Galactic open clusters like the Pleiades, h and $\chi$ Per, and Orion are well studied, not just because of their proximity but because of their rich stellar content. Berkeley 50 is thus interesting for being near another extreme, and possibly more typical of Galactic open clusters.

In Section~\ref{sec-obs}, we will discuss the observations and reductions, including the determination of photometry.    In Section~\ref{sec-mem} we will separate likely members from non-members. In Section~\ref{sec-HRD} we describe the classification of the stars for which we obtained spectra, and use this information, along with the photometry, to construct a physical H-R diagram for the cluster. We summarize our results in Section~\ref{sec-summary}.

\section{Observations and Reductions} \label{sec-obs}

We obtained both UBV imaging data and follow-up spectroscopy with the 4.3-meter LDT located at Happy Jack, Arizona.  We describe the observations and reductions in this section.

\subsection{Photometry}

Images were taken on the LDT using the Large Monolithic Imager (LMI).  The imager consists of a single 92.2-mm by 92.4-mm e2v CCD231-C6 chip with a multi-layer anti-reflective coating. It is operated at -120$^\circ$ C using a Stirling closed-cycle cooler.   The 125-mm Bonn shutter provides better than 1\% uniformity for exposures as short as 0.1~s, allowing broad-band photometry of even 10th magnitude stars in sub-arcsecond seeing.  Additional details can be found in the instrument manual.\footnote{http://www2.lowell.edu/users/massey/LMIdoc.pdf}  The CCD was binned 2$\times$2, resulting in a scale of 0\farcs24 per binned pixel.  After trimming, the resulting
field-of-view was 12\farcm3$\times$12\farcm3. The chip was read out through a single amplifier, with a gain of 2.89 electrons per analog-to-digital unit (ADU).
The device has a read-noise of 6.0 electrons.
The instrument is linear to better than 1\% up to a data value of 60,000 ADU (175,000 electrons) per unbinned pixel.

The images were kindly taken by Dr.\ Deidre Hunter prior to starting her own observing just as astronomical twilight ended on 2020 October 14, and consisted of a 0.5~s exposure in V, a 2~s exposure in B, and a 10~s exposure in U.  The observations were made with the cluster very near zenith, with an 
airmass of 1.00.  The image quality
was very good, with measured full-widths-at-half-maxima of 3.0 pixels (0\farcs7) at U, 2.7 pixels (0\farcs6) at B, and 2.0 pixels (0\farcs5) at V.

The instrumental signatures were removed using the {\sc ccdproc} routines in 
{\sc iraf}.\footnote{IRAF is distributed by the National Optical-Infrared Astronomy Research Laboratory, which is managed by the Association of Universities for Research in Astronomy (AURA) under a cooperative agreement with the National Science Foundation.}  For each frame, including calibration images, the over-scan region was used to find the bias offset value, and this number subtracted from the data on each frame.  The frames were then trimmed to 3069$\times$3076 pixels.   In order to remove the remaining two-dimensional structure due to the read-out, thirty bias-frames were (obtained at the start of the night) were averaged.  This master bias frame was subtracted from all of the data.  Flat-fielding was accomplished by combining five exposures obtained in twilight through each filter, with the telescope dithered between exposures to facilitate the removal of any stars via medianing.  The flats were normalized, and divided into the science data.

Standard stars selected from \cite{1992AJ....104..340L} were observed during evening twilight as part of  Dr.\ Hunter's main observing project, and included stars in three
Selected Areas (109, 110, 112), as well as components of PG1657+078. The
data ranged in airmass from 1.2 to 1.5, and with
colors from B-V=-0.15 to 2.33, providing good coverage. 
Aperture photometry was performed on 15 well-exposed stars in these fields, using a radius of 15 pixels (3\farcs6).  The {\sc photcal} routines in {\sc iraf} were used to solve the 
transformation equations; we settled on 
$$u=1.179+0.562X-0.103(U-B)$$
$$b=-0.878+0.257X-0.034(B-V)$$
$$v=-0.867+0.193X+0.042(B-V),$$
where the lower case letters refers to the instrumental magnitudes, $X$ refers to the airmass, and the zero-points are based on assigning a magnitude of 25.0 to an object with a count rate of 1 ADU per second.  Note that in order to determine the standard magnitudes (denoted by capital letters) these equations had to be solved iteratively. The transformations were poorly determined at $U$ (with scatter of 0.07~mag) but acceptable at $B$ (0.02~mag) and $V$ (0.01~mag).  We eliminated the U-B from our analysis.

For the cluster observations, we performed photometry using the implementation of {\sc daophot} \citep{1987PASP...99..191S} within {\sc iraf}.  On each frame, we first performed aperture photometry (with a 3-pixel radius) on all the stars that were 4$\sigma$ above the noise.  We then selected 3-4 isolated stars to be used to define the point-spread-function (PSF).  This PSF was then used to fit on all the stars on the frame simultaneously.  After the scaled PSFs were subtracted from the image, a new search identified any stars that had been hidden by neighboring stars, and the PSF-fitting re-done.  The zero-points of the PSF were based on the 3-pixel aperture photometry, and so for each frame an aperture correction was determined from isolated stars in order to correct this to the 15-pixel radius used on the standard stars.
The corrected photometry was then converted to the standard UBV system using the above equations.

Finally, the x and y pixel positions of each star were converted to right ascension ($\alpha_{\rm 2000}$) and declination ($\delta_{\rm 2000}$) using the {\sc astrometry.net} package\citep{2010AJ....139.1782L} using 2MASS stars \citep{2003yCat.2246....0C} as the reference frame. The resulting coordinates are good to about 0\farcs2.
Our photometry of the 1145 stars are given in Table~\ref{tab:all}.

\begin{deluxetable}{c c c c c c c c c c}[htbp!]
\tablecaption{\label{tab:all} Stars Measured in our LMI Field}
\tablewidth{0pt}
\tablehead{
\colhead{Star} &
\colhead{Radius\tablenotemark{a}} &
\colhead{$\alpha_{\rm 2000}$} &
\colhead{$\delta_{\rm 2000}$} &
\colhead{V} &
\colhead{B-V} &
\colhead{$\pi$\tablenotemark{b}} &
\colhead{$\mu$\tablenotemark{b}} &
\colhead{$\mu_{\alpha}$\tablenotemark{b}} &
\colhead{$\mu_{\delta}$\tablenotemark{b}} \\  & \arcmin & & & &  & mas & mas yr$^{-1}$& mas yr$^{-1}$ & mas yr$^{-1}$ }
\startdata
Be50-1 &  5.57 &  20:09:34.35 &  34:58:36.2 &  10.60 &  2.13 &   0.9199 &   2.44 &  -1.44 &  -1.97 \\
Be50-2 &  6.71 &  20:09:42.33 &  34:52:46.1 &  11.43 &  0.44 &   1.8954 &   2.35 &   1.06 &  -2.09 \\
Be50-3 &  2.62 &  20:10:05.94 &  34:55:44.8 &  11.58 &  0.52 &   0.6251 &   2.92 &  -1.16 &  -2.67 \\
Be50-4 &  6.01 &  20:09:33.80 &  34:56:12.0 &  11.76 &  0.17 &   1.0933 &   2.37 &   2.35 &  -0.28 \\
Be50-5 &  6.34 &  20:09:37.91 &  35:02:17.9 &  11.86 &  0.29 &   0.4204 &   2.21 &  -1.05 &  -1.94 \\
Be50-6 &  1.25 &  20:10:02.70 &  34:59:24.7 &  11.85 &  0.24 &   0.7265 &   2.42 &  -0.72 &  -2.30 \\
Be50-7 &  5.93 &  20:09:43.48 &  34:53:33.7 &  11.46 &  0.57 &   2.1981 &  11.65 &   2.75 &  11.32 \\
Be50-8 &  4.03 &  20:09:56.31 &  35:02:04.8 &  12.12 &  0.28 &   0.3910 &   5.52 &  -2.21 &  -5.05 \\
Be50-9 &  5.47 &  20:10:28.06 &  34:58:19.9 &  12.25 &  0.35 &   0.5427 &   4.52 &  -2.67 &  -3.64 \\
Be50-10 &  2.79 &  20:10:00.32 &  34:55:25.0 &  12.03 &  0.36 &   1.4324 &   4.32 &   3.93 &   1.81 \\
\enddata
\tablecomments{Table 1 is published in its entirety in the machine-readable format.}
\tablenotetext{a}{Angular separation from the RSG.}
\tablenotetext{b}{Data from Gaia DR3 (Gaia Collaboration et al.\ 2023).}
\end{deluxetable}

\subsection{Spectroscopy}
\label{sec-spec}
Prior to the imaging data being obtained, a spectrum of the alleged RSG was taken on 2020 September 28 by PM and our colleague Dr.\ Kathryn Neugent; the data showed TiO bands characteristic of a K5-7 star. Spectra of a few of the brightest neighboring stars were taken on 2021 October 22; these were all characteristic of B-type dwarfs, consistent with what we would expect for a sparse cluster containing a single RSG.  The majority of our spectra were taken on 2022 October 5-6; including obtaining better signal-to-noise data on all stars previously observed. The sample was based on our photometry and a casual inspection of the DR3 proper motions \citep{gaia,gaiadr3}. 

All the spectroscopy was carried out on the LDT using the DeVeny optical spectrograph.  The detector is a 2048$\times$512 e2v CCD42-10 deep depletion device with 13.5~$\mu$m pixels.  It is operated at a gain of 1.5 electrons ADU$^{-1}$, and has a full-well of about 100,000 electrons. A Stirling closed-cycle cooler chills the chip to $-110^\circ$ C. A slit viewing camera is used to position stars on the slit. The spatial scale on the LDT is 0\farcs34 pixel$^{-1}$.

A 600 line mm$^{-1}$ grating blazed at 4900~\AA\ was used with a grating tilt of 26\fdg95, which provided wavelength coverage from 3700-5950~\AA\ and a dispersion of 1.12~\AA~pixel$^{-1}$.
The slit was opened to a width of 1\farcs0, which provided a spectral resolution of 2.5~\AA.  Exposure times of 300-1800 s were used, depending upon the brightness of the object; longer exposures were broken up into three pieces in order to more easily remove radiation events (``cosmic rays").  Spectrophotometric standards were observed at the start of each night to allow flux calibration. All observations were made by rotating the spectrograph so the slit was oriented to the parallactic angle. Wavelength calibration was provided by an exposure of Hg-Cd-Ar comparison lamps at the start of the night. The spectrograph suffers from some flexure, but, unfortunately, the procedure for taking comparison exposures is time-consuming, making it impractical to take comparison spectra for each object. Thus, our data are not good enough for accurate radial velocity information. Dome flats and bias frames were taken in the afternoon.

Reduction of the data took place using the standard {\sc iraf} routines.  First, the overscan was measured and subtracted from each image. Next, images were trimmed to 2033$\times$375 pixels.   The averaged bias frame was subtracted from each exposure to remove any residual two-dimensional structure to the bias.  The flat-field exposures were averaged, normalized, and divided into the science frames.

A $\pm$4 pixel extraction aperture was defined around the peak of the star. The position of the stellar spectrum was automatically measured as a function of wavelength, and fit with a third-order cubic spline function. This trace was used to provide a star-specific extraction of the afternoon's comparison spectrum, and a wavelength solution computed using another third-order cubic spline, leaving typical residuals of 0.02~\AA. The trace was then used to perform an optimized extraction, with sky values interpolated from either side of the stellar spectrum, and the wavelength calibration applied.

\section{Membership}
\label{sec-mem}

In order to establish membership, we relied upon data from Gaia DR3 \citep{gaiadr3}. From the brightness of the RSG and the B-type stars, we knew that the cluster was at a distance of several kiloparsecs. At this distance, Gaia parallaxes are sufficiently imprecise that they were not suitable for a first pass, but we could rely upon Gaia proper motions to eliminate unlikely members. (Once we restrict the sample by proper motions we will use parallaxes to refine membership.) 
Table \ref{tab:all} includes the Gaia proper motions and parallaxes for the complete sample of 1145 stars we measured in our 12\farcm3$\times$12\farcm3 LMI frames.

In order to establish the likely proper motions of members, we began by considering only those stars within a 5\arcmin\  radius of the RSG.  The median proper motion of this sample is 6.11 mas yr$^{-1}$.
Within this sample, the median proper motion in right ascension and declination to be $\mu_{\alpha}$ = -3.32 mas yr$^{-1}$ and $\mu_{\delta}$ = -5.13 mas yr$^{-1}$ respectively. (These values agree well the statistical determination of the proper motions by \citealt{2023A&A...673A.114H} who found $\mu_{\alpha}$ = -3.37 mas yr$^{-1}$ and $\mu_{\delta}$ = -5.13 mas yr$^{-1}$.)  Based on the distribution of values around these  values, we chose to restrict the sample to those stars whose proper motions were within 3$\sigma$ of these median values, based on the tabulated Gaia uncertainties. Implementing these proper motion restrictions eliminated $\sim$86\% of the stars, and reduced the sample of potential members to 163.  

In the next section, we will describe our spectroscopy of the brighter stars.  With the exception of the RSG, these were all dwarfs with spectral types of mid-B through A0. A comparison of the Gaia parallaxes for these stars established a median distance of 3.8~kpc for the cluster using the values tabulated by \citet{2021AJ....161..147B}. (We note that this distance agrees with the 3.8~kpc value  found by \citealt{2023A&A...673A.114H}, who used a statistical approach to the Gaia DR3 data without the benefit of spectroscopy to help limit membership.)  An inspection of the distribution of parallaxes of these proper-motion selected sample allowed us to impose a cutoff of 0.215 mas to 0.265 mas for the parallaxes of potential members. This reduced our sample of likely cluster members to 64 stars. (Of course, some stars lacked parallax information, and thus this sample should not be considered to be a magnitude-limited sample.)

\begin{figure}
\epsscale{0.45}
\plotone{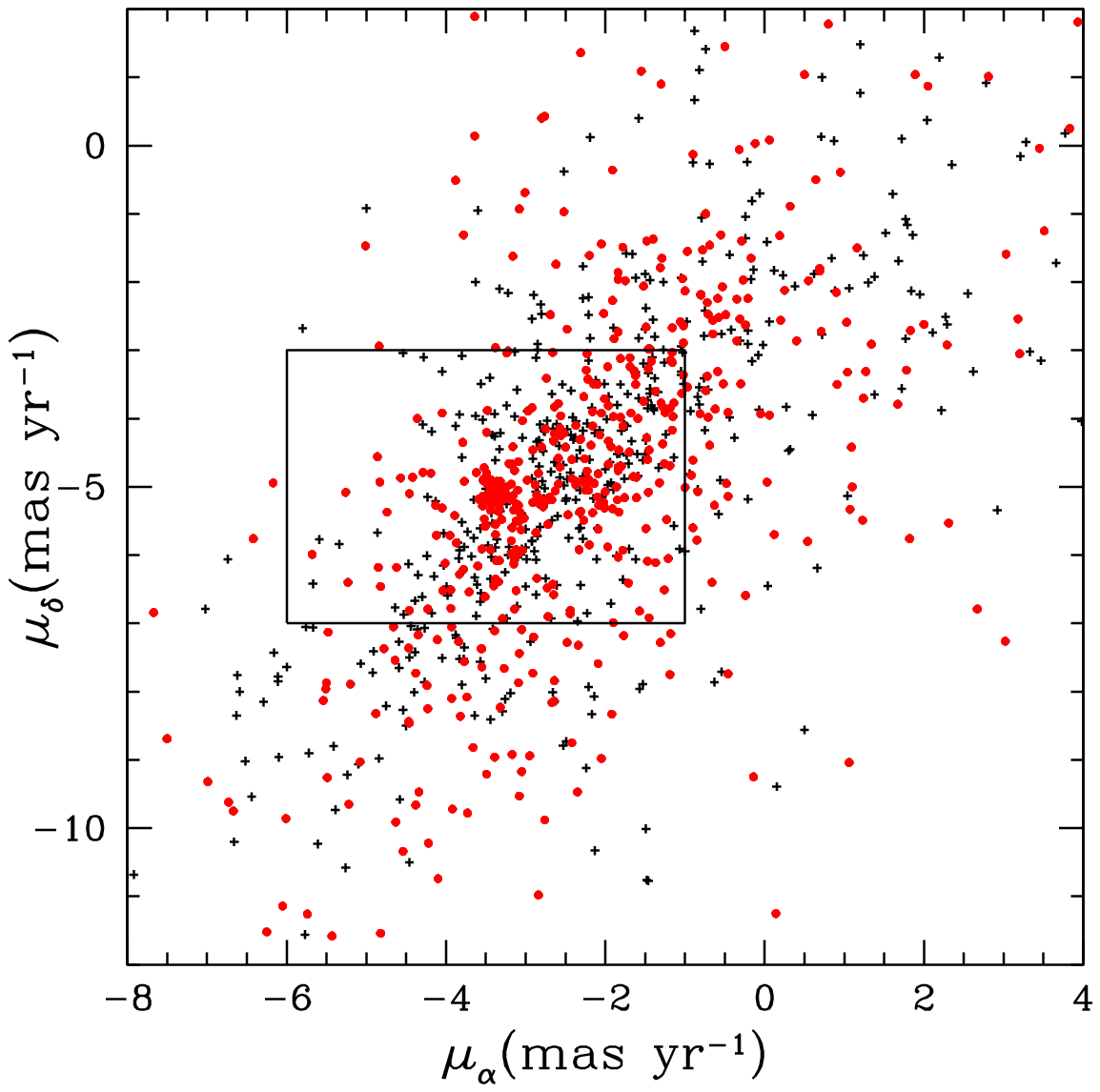}
\plotone{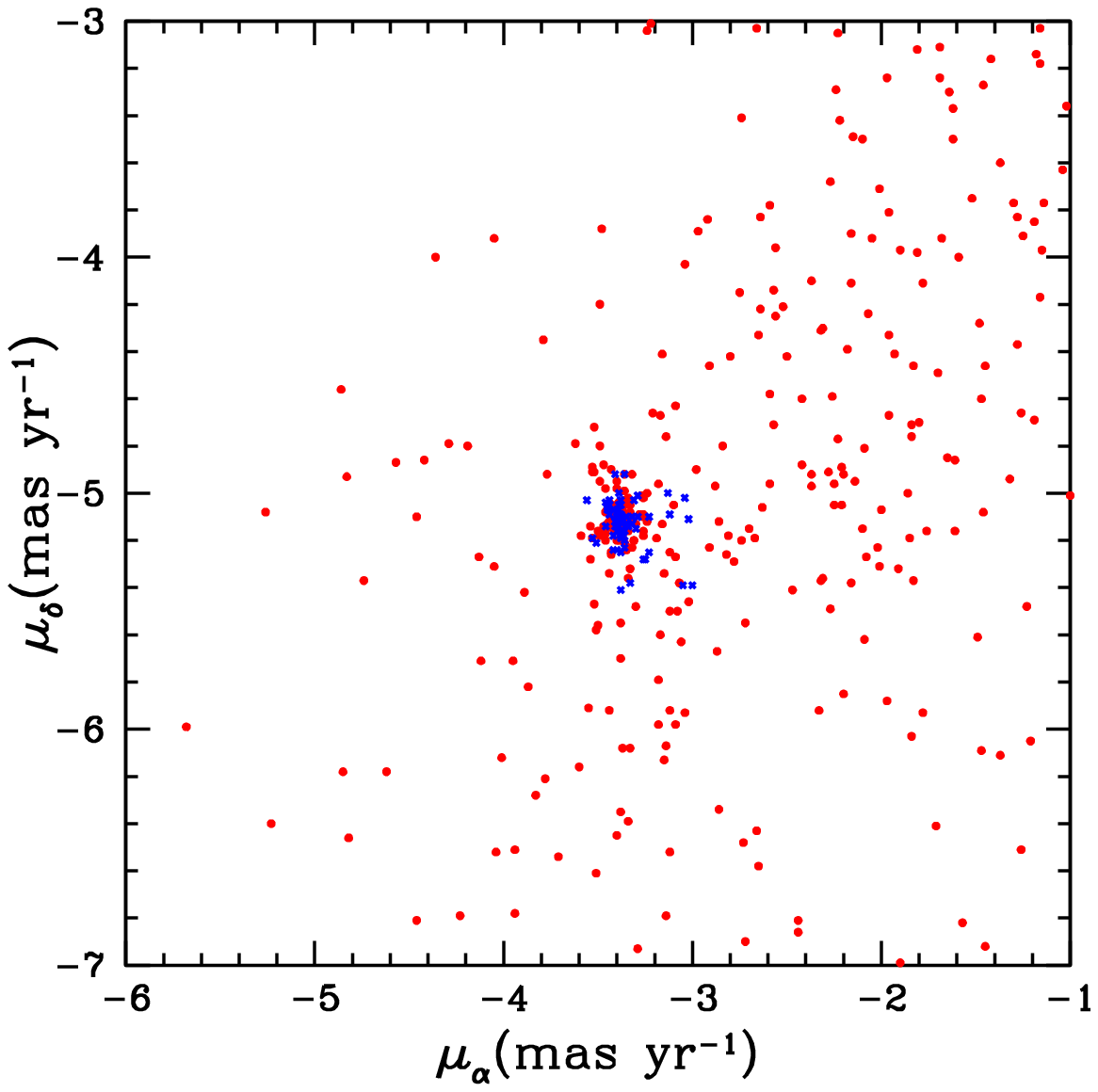}
\caption{\label{fig:pms}.  Selection of cluster members.  {\it Left:} The proper motions of stars in the LMI field are shown by black + signs.  The red circles denote the proper motions of stars within 5\arcmin\ of the center of the cluster. Note the clump of red points near the center; these were the potential cluster members.  {\it Right:} Red circles again show the proper motions of stars within 5\arcmin\ of the center of the cluster, while the blue x's denote the proper motions of the 64 cluster members selected after applying the additional parallax criteria $0.215$~mas $\le \pi \le 0.265$~mas.}
\end{figure}

This process is illustrated in Figure~\ref{fig:pms}, where we show on the left the distribution of proper motions over the field along with those in the central 5\arcmin. We see a clumping near $\mu_\alpha -3.3$ and $\mu_\delta=-5.1$. We show the expansion of the boxed region on the right, where the blue x's show the final sample of cluster members after we have imposed parallax restrictions.  A few of the members have distances that are slightly over 5\arcmin\ from the RSG.

 As we noted earlier, the RSG has a parallax outside this range, but its ``renormalized unit weight error" suggests that its value is uncertain.  This is a commonly an issue for stars that are unresolved binaries; as we will see in the next section, we find that the spectrum of our RSG indicates a binary companion. 

We identify these 64 cluster members in Table~\ref{tab:gmem}, along with their physical properties that we derive in subsequent sections.

\begin{deluxetable}{l l c c c c c}[htbp!]
\tablecaption{\label{tab:gmem} Cluster Members}
\label{tab:fmem}
\tablewidth{0pt}
\tablehead{
\colhead{Star} &
\colhead{Sp.Type} &
\colhead{V} &
\colhead{B-V} &
\colhead{$M_{V}$ } &
\colhead{ log $T_{\rm eff}$} &
\colhead{log($L/L_{\odot}$)} }
\startdata
Be50-63 &     K5-7 I + B &  11.96 &  2.06 & -3.73 &      3.580 &       3.80 \\
Be50-50 &           B5 V &  13.71 &  0.88 & -1.98 &      4.182 &       3.30 \\
Be50-47 &           B8 V &  13.71 &  0.75 & -1.98 &      4.057 &       3.10 \\
Be50-52 &           B8 V &  13.94 &  0.67 & -1.75 &      4.057 &       3.01 \\
Be50-61 &           B5 V &  14.06 &  0.77 & -1.63 &      4.182 &       3.16 \\
Be50-67 &           B5 V &  14.40 &  0.74 & -1.29 &      4.182 &       3.02 \\
Be50-69 &           B5 V &  14.51 &  0.68 & -1.17 &      4.182 &       2.98 \\
Be50-71 &           B5 V &  14.58 &  0.63 & -1.10 &      4.182 &       2.95 \\
Be50-85 &           B8 V &  14.66 &  0.77 & -1.03 &      4.057 &       2.72 \\
Be50-92 &           B5 V &  14.76 &  0.74 & -0.93 &      4.182 &       2.88 \\
Be50-96 &           B5 V &  14.78 &  0.69 & -0.91 &      4.182 &       2.87 \\
Be50-107 &           A0 V &  14.89 &  0.74 & -0.80 &      3.990 &       2.32 \\
Be50-132 &           B8 V &  15.12 &  0.83 & -0.57 &      4.057 &       2.54 \\
Be50-139 &           B8 V &  15.20 &  0.84 & -0.49 &      4.057 &       2.50 \\
Be50-123 &           B8 V &  15.22 &  0.68 & -0.47 &      4.057 &       2.50 \\
Be50-147 &           B5 V &  15.23 &  0.86 & -0.46 &      4.182 &       2.69 \\
Be50-137 &           B5 V &  15.31 &  0.76 & -0.38 &      4.182 &       2.66 \\
Be50-180 &         \nodata &  15.62 &  0.83 & -0.07 &      4.061 &       2.19 \\
Be50-171 &         \nodata &  15.66 &  0.74 & -0.03 &      4.184 &       2.47 \\
Be50-196 &         \nodata &  15.76 &  0.86 &  0.07 &      4.023 &       2.03 \\
\enddata
\tablecomments{Table 2 is published in its entirety in the machine-readable format.}
\end{deluxetable}


\section{Constructing the HRD}
\label{sec-HRD}
\subsection{Spectral Classification}

The spectra allow us to assign more accurate temperatures to the hot stars than we can accomplish by photometry alone, and in addition they provide the means for making an accurate accessment of the color-excess and hence extinction, needed for computing luminosities. We begin by classifying the spectra after normalizing them to the continuum.  

With the exception of the RSG, all of the bright cluster members we observed spectroscopically were of mid-to-late B-type plus one A0 star. For the classification, we followed the criteria illustrated in \citet{1990PASP..102..379W} and \citet{2009ssc..book.....G}.
The classification of B stars depends primarily on the relative strengths of the Si\,{\sc iv} $\lambda4089$, Si\,{\sc iii} $\lambda$4553-7 complex, and Si\,{\sc ii} $\lambda$4128 line, with earlier B-types having greater strengths of the higher ionization Si lines.  Specifically, a B0 star has Si\,{\sc iv} much stronger than Si\,{\sc iii} and still shows weak He\,{\sc ii}~$\lambda$4686.
At B0.5 the Si\,{\sc iv} and Si\,{\sc iii} lines are of equal strength. By B1 Si\,{\sc iv} is weaker than Si\,{\sc iii}. At B2, Si\,{\sc iv} is gone and Si\,{\sc iii} is still much stronger than Si\,{\sc ii}. By B5, Si\,{\sc iii} and S\,{\sc ii} are of equal strength.  At B8, S\,{\sc ii} dominates over Si\,{\sc iii} and the Mg\,{\sc ii} $\lambda4481$ rivals He\,{\sc i}$\lambda$4471 in strength.  This is all well illustrated in Figures 4.1 and 4.2 of \citet{2009ssc..book.....G}. At A0 the He\,{\sc i} lines are gone, with the spectral subtype determined by the relative strengths of the Ca\,{\sc ii} K line relative to H$\epsilon$ \citep{2009ssc..book.....G}.  A complication is that the absolute strengths of the metal lines are dependent on surface gravity (i.e., luminosity class), and are weak in dwarfs, requiring good signal-to-noise to ascertain the exact sub-type.  The weakness of the metal lines, and the widths
of the hydrogen Balmer lines, confirms that all of these stars are dwarfs.

\begin{figure}[h!]
  \centering 
  \includegraphics[scale=0.8]{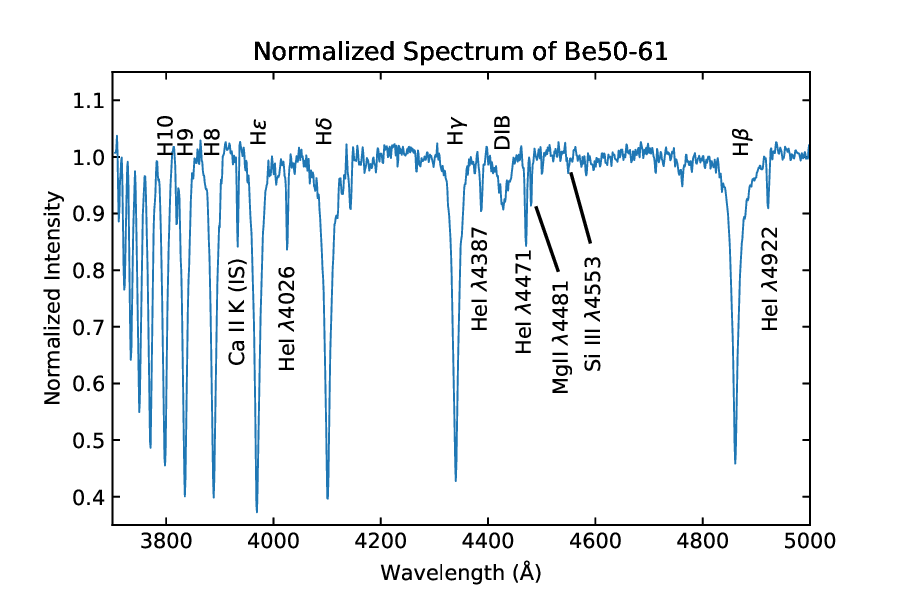}
  \caption{The normalized spectrum of star Be50-61. We show only the classical classification region, 3700-5000\AA. This star was classified as a B5~V. We have labeled the principal lines.  The DIB is the diffuse interstellar band at 4430\AA.}
\label{fig-norm}
\end{figure}

In Table~\ref{tab:gspect}, we list the spectral classifications. Figure~\ref{fig-norm} displays one of the normalized spectra,
a B5~V, illustrating the quality of our spectra.

The RSG was classified based using the strengths of the TiO molecular bands, following \citet{2017ars..book.....L} and references therein. The 5167~\AA\ band is obvious, but
the 4954~\AA\ band is weak or absent, making the spectral type K5-7.  
A careful inspection of the spectrum reveals the upper Balmer lines in absorption, a sign that the star is a RSG+B
binary. Although recent work has identified hundreds of such RSG+B systems in the Magellanic Clouds, M31, and M33 \citep{2019ApJ...875..124N,2020ApJ...900..118N,2021ApJ...908...87N}, only about a dozen of such systems are known in the Milky Way (\citealt{2018AJ....156..225N} and references therein). Evolutionary theory shows that the companions of RSGs are invariably of B-type \citep{2018AJ....156..225N,2019ApJ...875..124N}, and indeed that is what we find here.

Based on the assumption of membership, the star must be a supergiant as we argue in the next section.  But we can we find spectroscopic evidence of its luminosity class?  The best luminosity criteria for RSGs are in the far red.  We have examined the very low resolution Gaia spectrum of the star, which extends out to 1$\mu$m.  The strong jump from 8500\AA\ to 9000\AA\ for stars in this temperature range indeed suggests a supergiant luminosity class, but not as strong as a Ia, say, where the jump would be stronger.  

\begin{figure}[ht!]
  \centering
  \begin{minipage}[b]{0.49\textwidth}
    \centering
    \includegraphics[width=\textwidth]{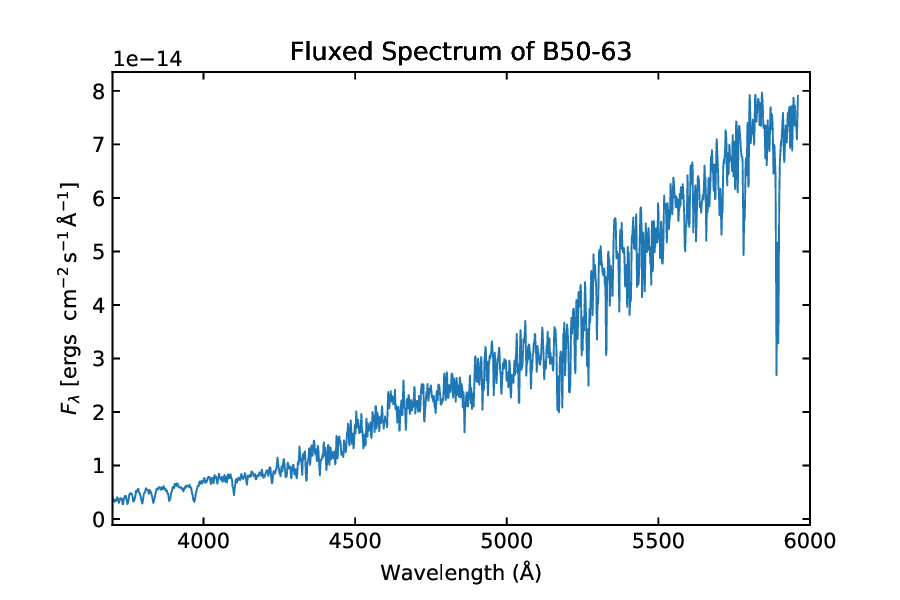}

  \end{minipage}
  \hfill
  \begin{minipage}[b]{0.49\textwidth}
    \centering
    \includegraphics[width=\textwidth]{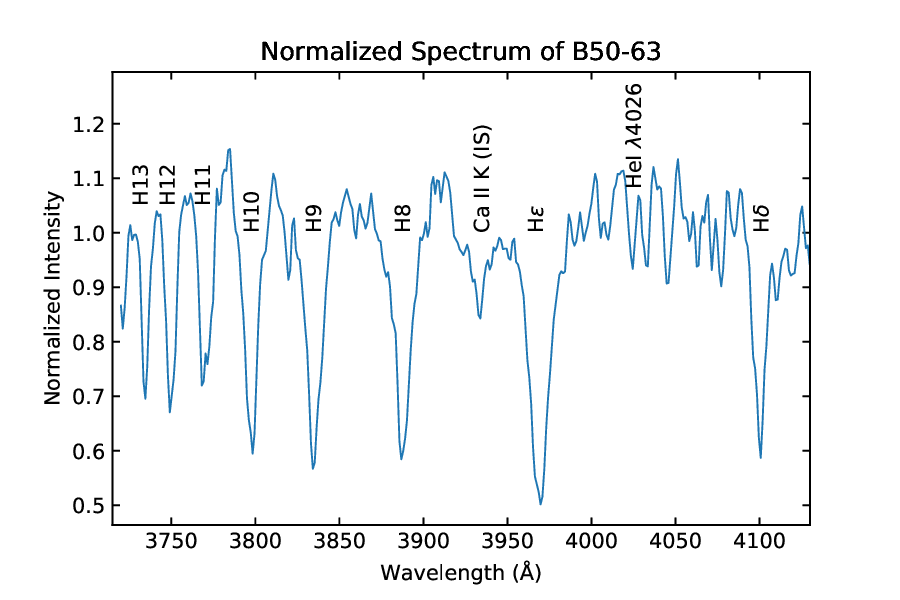}
  \end{minipage}
  \caption{Spectrum of Be50-63, the RSG+B star. {\it Left} The fluxed spectrum shows the TiO $\lambda 5167$ band but no others, making the spectral type of the RSG component K5-7.  {\it Right:} The near-UV shows the upper Balmer lines, characteristic of the presence of a B-type companion.}
  \label{flux}
\end{figure}

\begin{deluxetable}{l c c c c c c c c }
\tablecaption{\label{tab:gspect} Spectral Types and Adopted Quantities}
\tablewidth{0pt}
\label{tab:smem}
\tablehead{
\colhead{Star} & 
\colhead{Type} &
\colhead{B-V} &
\colhead{$(B-V)_{0}^{\rm type}$} &
\colhead{$E(B-V)_{\rm type}$} &
\colhead{V} &
\colhead{$M_{V}$ } & 
\colhead{log($T_{\rm eff}$)} &
\colhead{log($L/L_{\odot}$)} }

\startdata
Be50-50 &           B5 V &  0.88 &   -0.17 &    1.05 &  13.71 & -1.98 &      4.182 &       3.30 \\
Be50-71 &           B5 V &  0.63 &   -0.17 &    0.80 &  14.58 & -1.10 &      4.182 &       2.95 \\
Be50-137 &           B5 V &  0.76 &   -0.17 &    0.93 &  15.31 & -0.38 &      4.182 &       2.66 \\
Be50-147 &           B5 V &  0.86 &   -0.17 &    1.03 &  15.23 & -0.46 &      4.182 &       2.69 \\
Be50-61 &           B5 V &  0.77 &   -0.17 &    0.94 &  14.06 & -1.63 &      4.182 &       3.16 \\
Be50-67 &           B5 V &  0.74 &   -0.17 &    0.91 &  14.40 & -1.29 &      4.182 &       3.02 \\
Be50-92 &           B5 V &  0.74 &   -0.17 &    0.91 &  14.76 & -0.93 &      4.182 &       2.88 \\
Be50-96 &           B5 V &  0.69 &   -0.17 &    0.86 &  14.78 & -0.91 &      4.182 &       2.87 \\
Be50-69 &           B5 V &  0.68 &   -0.17 &    0.85 &  14.51 & -1.17 &      4.182 &       2.98 \\
Be50-139 &           B8 V &  0.84 &   -0.11 &    0.94 &  15.20 & -0.49 &      4.057 &       2.50 \\
Be50-132 &           B8 V &  0.83 &   -0.11 &    0.94 &  15.12 & -0.57 &      4.057 &       2.54 \\
Be50-85 &           B8 V &  0.77 &   -0.11 &    0.88 &  14.66 & -1.03 &      4.057 &       2.72 \\
Be50-47 &           B8 V &  0.75 &   -0.11 &    0.86 &  13.71 & -1.98 &      4.057 &       3.10 \\
Be50-123 &           B8 V &  0.68 &   -0.11 &    0.79 &  15.22 & -0.47 &      4.057 &       2.50 \\
Be50-52 &           B8 V &  0.67 &   -0.11 &    0.78 &  13.94 & -1.75 &      4.057 &       3.01 \\
Be50-107 &           A0 V &  0.74 &   -0.02 &    0.76 &  14.89 & -0.80 &      3.990 &       2.32 \\
Be50-63 &     K5-7~I+B &  2.06 &     \nodata &   \nodata &  11.96 & -3.73 &      3.580 &       3.78 \\
\enddata
\tablecomments{The adopted $(B-V)_o^{type}$ values used to compute the individual $E(B-V)$ values come from Table 15.7 of \citet{2000asqu.book.....C}, and do the effective temperatures based on the spectral types.}
\end{deluxetable}

\subsection{Extinction}

The spectral types allow us to determine the reddening of stars in the cluster, allowing us to compute luminosities since the distance is known.  We adopt the intrinsic $(B-V)_o$ colors as a function of spectral type given in Table 15.7 of \citet{2000asqu.book.....C}.  These values are in good agreement with  \citet{1970A&A.....4..234F} values for B stars.
The average $E(B-V)$ color excess we find is 0.90 mag. There was no evidence for spatial variations in the reddening.  We adopt an extinction correction $A_V=2.79$~mag, assuming a ratio $R_V$ of total-to-selective extinction of 3.1.  Thus, adopting the Gaia-derived distance of 3.8~kpc (corresponding to a true distance modulus of 12.9) leads to an apparent distance modulus of $V-M_V=15.7.$

\subsection{Assigning physical properties}
For the main-sequence stars with spectral information, we assign effective temperatures using the calibrations
given in Table 15.7 of \cite{2000asqu.book.....C}.  We note that the connection between spectral subtypes and temperatures could be
better refined using modern stellar atmosphere models.  However, the adopted calibration agrees
well with the recent fundamental study by \citet{2019ApJ...873...91G}.   

For stars lacking spectroscopy,
we used the de-reddened colors to assign temperatures, applying the relationship derived by
\citet{2002ApJ...576..880S}:
$$\log T_{\rm eff}  = 3.9889 - 0.7950(B-V)_{0} + 2.1269 (B-V)_{0}^2 -3.9330(B-V)_{0}^3$$
$$ + 3.5860(B-V)_{0}^4-1.5531(B-V)_{0}^5 + 0.2544(B-V)_{0}^6,$$
where we have de-reddened our observed $B-V$ values by the color excess $E(B-V)=0.90$; i.e.,
$(B-V)_0=(B-V)-0.90$.  The \citet{2002ApJ...576..880S} effective temperature relation came about from
a combination of observed \citep{1996ApJ...469..355F} and theoretical \citep{1992IAUS..149..225K} colors and effective temperatures.

These effective temperatures were then used with our adopted distance to derive the
bolometric luminosities.   With the exception of the RSG, our stars span the temperature range $\log T_{\rm eff} = 3.75$ to 4.25~dex.  To determine the bolometric corrections (BCs), we again adopt the work of \citet{2002ApJ...576..880S} (based primarily on \citealt{1997AJ....113.1733H}), which used the following relations:
$${\rm BC}=-8.58 +8.4647\log T_{\rm eff}-1.6125(\log T_{\rm eff})^2     {\phantom{Y} \rm for\phantom{1}} \log T_{\rm eff}>4.1$$
$${\rm BC}=-312.90+161.466\log T_{\rm eff}-20.827(\log T_{\rm eff})^2 {\phantom{Y} \rm for\phantom{1}} 4.1>\log T_{\rm eff}>3.83 \phantom{Y} {\rm and}$$
$${\rm BC}-346.82+182.396\log T_{\rm eff}-23.981(\log T_{\rm eff})^2 {\phantom{Y} \rm for\phantom{1}} 3.83>\log T_{\rm eff}>3.55.$$

This allowed us to compute the bolometric magnitudes $M_{\rm bol}=V-A_V-12.9+{\rm BC}$.  The bolometric
luminosity then follows from adopting the usual 4.75 mag value for the bolometric magnitude of the sun; i.e.,
$\log L/L_\odot = (M_{\rm bol} - 4.75)/-2.5$.

We placed the RSG on the  HRD assuming that the B companion
contributes negligibly to the V-band brightness (see, e.g., \citealt{2018AJ....156..225N}).  The effective
temperature of a  K5-7 I Galactic star is 3840 K ($\log T_{\rm eff}=3.58$) with a bolometric correction of $-1.16$~mag at $V$,
according to \citet{2005ApJ...628..973L}.  Correcting V=11.96 by $A_V=2.8$~mag, and
applying a true distance modulus of 12.90 mags (3.8~kpc) leads to $M_V=-3.74$ and a luminosity of $\log L/L_\odot$=3.86.
We note that this absolute magnitude is more consistent with a fainter Ib, but far more luminous than a than a K5-7 giant, where $M_V\sim0$ (see, e.g., Table 15.7 in \citealt{2000asqu.book.....C}).

We can check the RSG's luminosity by using its near-IR 2MASS photometry.  This should be even less
affected by the hot companion than the optical.   The K-band brightness is 5.38~mag.  An $A_V=2.79$~mag translates to $A_K=0.33$~mag.  Thus, $M_K=-7.4$.  The K-band bolometric correction
for a K5-7 star is +2.7~mag and hence the luminosity we derive is $\log L/L_\odot=3.78$.
The agreement with the optically derived value is excellent; i.e., $\log L/L_\odot\sim$3.8-3.9.

We include the adopted values for $\log T_{\rm eff}$ and $\log L/L_\odot$ in Tables~\ref{tab:gmem} and \ref{tab:gspect}.

\subsection{Comparison with Evolutionary Tracks}

\begin{figure}[h!]
\epsscale{1.2}
\plotone{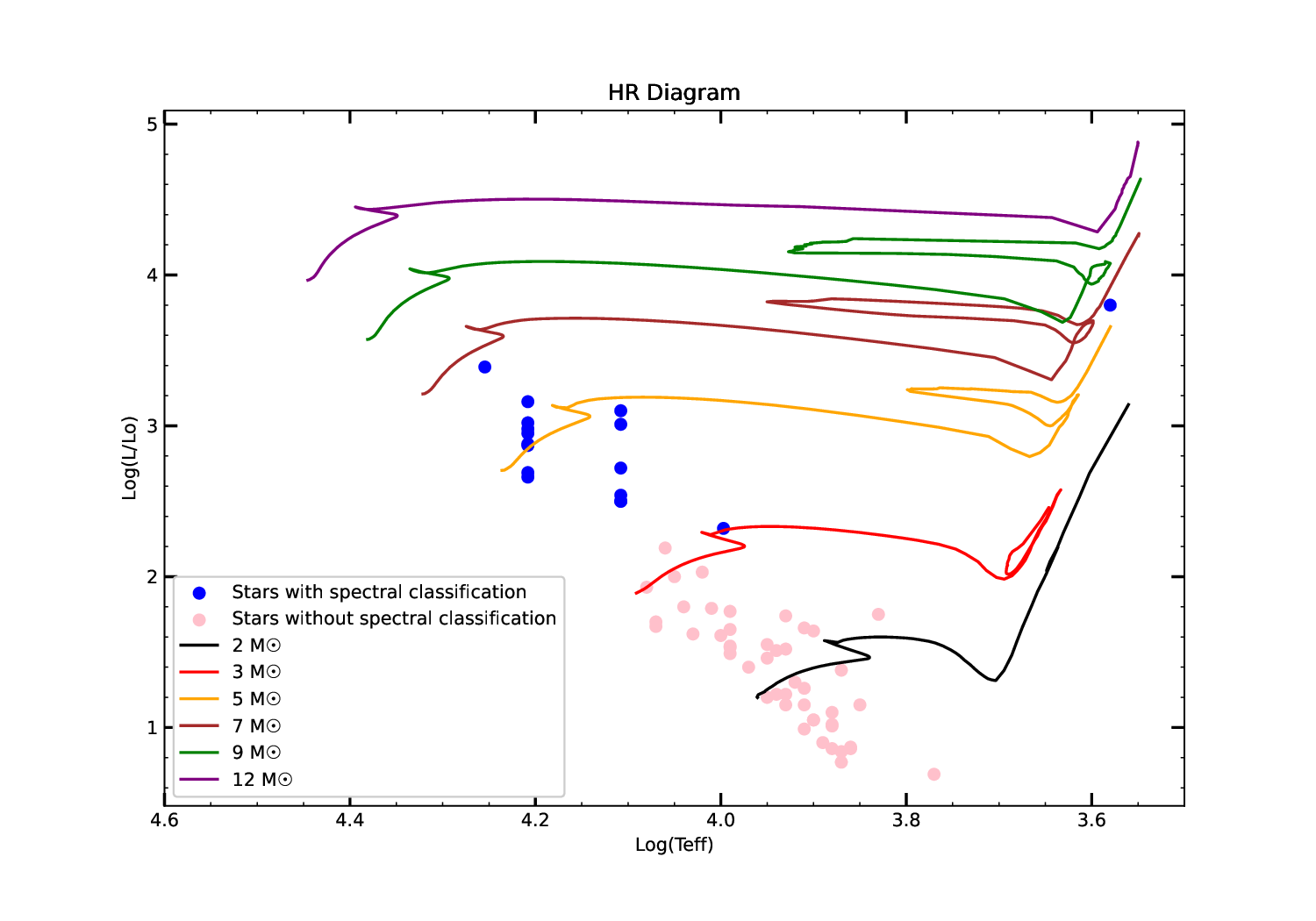}
  \caption{The H-R diagram for Berkeley 50. The blue points have been placed based on their spectral types; the light pink points by their photometry. The colored lines correspond to the Geneva evolutionary tracks computed with rotation \citep{2012A&A...537A.146E}.  The dashed curve corresponds to a 60$M_\odot$ isochrone.}  
\label{fig-HRD}
\end{figure}

In Figure~\ref{fig-HRD} we plot our H-R diagram, along with the Geneva evolutionary tracks of 
 \cite{2012A&A...537A.146E}. The tracks
correspond to a solar-like metallicity of $Z=0.014$, and assume that stars reach the zero-age-main-sequence with initial rotation of 40\% of their breakup speed.

We see that our work defines a strong main-sequence for Berkeley 50. The most massive stars correspond to $\sim 7 M_\odot$.  The RSG falls exactly where one would then expect, consistent with the most massive main-sequence star.  The corresponding age from the \citet{2012A&A...537A.146E} Geneva evolutionary tracks is 59~Myr using the tracks
that include rotation. Using the non-rotating tracks would imply an age of 50~Myr.   We conclude
that the age of the cluster is 50-60~Myr.

The uncertainties in our HRD are hard to quantify.  We note that the classification of B dwarfs is tricky given that
the criteria depend upon the relative strengths of metal lines which are quite weak in dwarfs.  We estimate an uncertainty of $+/-$ 2 spectral subclasses; i.e., a star we call B8~V could be as early as B5.
This introduces an uncertainty of 0.1~dex in $\log T_{\rm eff}$ and 0.4~dex in $\log L/L_\odot$. 

The closest analog we can find to Berkeley 50 is the cluster Messier 6 (NGC 6405).  It has a similar age, and contains a lone RSG, BM Sco, an MK standard for the K2~Ib (see, e.g., \citealt{1965PASP...77...19T, 1975AJ.....80...11V,2018AJ....156..121G} and \citealt{2021MNRAS.504..356D}),
although Messier 6 is far more populous than Berkeley 50.

\section{Summary} 
\label{sec-summary}
In this study of Berkeley 50, we sought to understand the stellar content of this Galactic open cluster. We were first interested in the cluster due to the apparent presence of a RSG in the cluster, and after spectroscopic analysis, we determined it is a K5-7 I  with a B-type companion. We obtained UBV photometry of 1145 stars in a 12\farcm3 $\times$ 12\farcm3 area. To determine  potential membership, we used Gaia proper motions near the RSG to identify stars suitable for spectroscopy.

We obtained optical spectroscopy  of the 17 brightest stars near the RSG, and following the criteria in \citet{1990PASP..102..379W}, we classified the stars as mid-to-late B-type stars, with two exceptions, the RSG and one A0 star. The proper motions and parallaxes of these stars were all similar, which allowed us to establish  64 stars as certain members of this cluster, although we recognize this is not a complete census. This cluster may correspond more to typical  Galactic open clusters than do the well-populated clusters that have been well observed.

Using a comparison of the Gaia parallaxes for the spectroscopically confirmed members, we determined a distance of 3.8 kpc for the cluster, and our spectra showed that the reddening was uniform with E(B-V)=0.9. We assigned temperatures and bolometric corrections for the complete sample, using spectra when available, and photometry when not. For stars without spectroscopy, we used the de-reddened colors to assign temperature, using the relationship derived by \citet{2002ApJ...576..880S}. Using the effective temperatures of the cluster members with and without spectra data, we used our adopted distance of 3.8 kpc and derived the bolometric luminosities, applying \citet{2002ApJ...576..880S} again.

We compiled an HR diagram using these data and compared the stars’ locations to the Geneva evolutionary tracks from \cite{2012A&A...537A.146E}. The HR diagram depicts a strong main sequence for the cluster, except for the RSG, the most massive star in the cluster corresponding to  $\sim 7 M_\odot$. Comparison with the Geneva evolutionary tracks sets the cluster's age at 50-60 Myr.

The recent study by \citet{2023A&A...673A.114H} adopted an approximate  Bayesian neural network framework using variational inference in order to conduct a blind, all-sky search for open clusters using DR3.  Further, they
then computed extinction, distances, and even ages for their clusters.  We became aware of their work only after ours was finished, and we note, with admiration, that their statistical approach determined many of the same answers we found by our study.   Our extinction was based on spectroscopy and our own {\it BV} values; assuming $R_V=3.1$ we derive $A_V=2.79$~mag. Their statistical approach determined $A_V=2.78-3.18$~mag. We determined a distance using Gaia to determine membership based on parallaxes and proper motions, with our optical ground-based  spectroscopy used to anchor the distributions by identifying certain members.  Our value of 3.8~kpc is the same as what they derived. Our age estimate of 50-60~Myr was determined by physical properties (luminosity and temperature) of the RSG with comparison with stellar evolutionary tracks; their approach identified the age as 35-89~Myr.  That said, their study missed including the RSG, arguably the most interesting member of the cluster.

\begin{acknowledgments}

Lowell Observatory sits at the base of mountains sacred to tribes throughout the region. We honor their past, present, and future generations, who have lived here for millennia and will forever call this place home. 

M.S.'s participation was partially supported by the NASA Space Grant program at Northern Arizona University.  Additional support was provided by Lowell Observatory through the Bill Goddard Fund,  BF Foundation funds, and the Young Scholars Fund.  Dr.\ Kathryn Neugent helped with the overall design of the project, and was involved in the preliminary spectroscopy. Dr.\ Deidre Hunter generously obtained the imaging data used in this project, as well as the standard star solutions, and provided much moral support to P. M. throughout the completion of this manuscript.  We thank the anonymous referee for useful comments which improved this manuscript.

These results made use of the Lowell Discovery Telescope (LDT) at Lowell Observatory.  Lowell is a private, non-profit institution dedicated to astrophysical research and public appreciation of astronomy and operates the LDT in partnership with Boston University, the University of Maryland, the University of Toledo, Northern Arizona University and Yale University. The Large Monolithic Imager was built by Lowell Observatory using funds provided by the National Science Foundation (AST-1005313)

This work has also made use of data from the European Space Agency (ESA) mission
{\it Gaia} (\url{https://www.cosmos.esa.int/gaia}), processed by the {\it Gaia}
Data Processing and Analysis Consortium (DPAC,
\url{https://www.cosmos.esa.int/web/gaia/dpac/consortium}). Funding for the DPAC
has been provided by national institutions, in particular the institutions
participating in the {\it Gaia} Multilateral Agreement.

\end{acknowledgments}

\facilities{LDT (LMI, DeVeny), Gaia, FLWO:2MASS}


\bibliographystyle{aasjournal}
\bibliography{references}{}

\end{document}